# Optimization of Magnetic Refrigerators by Tuning the Heat Transfer Medium and Operating Conditions

Mohammadreza Ghahremani[1,2], Amir Aslani[2], Lawrence H. Bennett[2], *Senior Member, IEEE,* and Edward Della Torre[2], *Fellow, IEEE*

[1]Dept. of Computer Science, Mathematics, and Engineering, Shepherd University, Shepherdstown WV 25443
[2]Dept. of Electrical and Computer Engineering, The George Washington University, Washington DC 20052

A new experimental test bed has been designed, built, and tested to evaluate the effect of the system's parameters on a reciprocating Active Magnetic Regenerator (AMR) near room temperature. Bulk gadolinium was used as the refrigerant, silicon oil as the heat transfer medium, and a magnetic field of 1.3 T was cycled. This study focuses on the methodology of single stage AMR operation conditions to get a higher temperature span near room temperature. Herein, the main objective is not to report the absolute maximum attainable temperature span seen in an AMR system, but rather to find the system's optimal operating conditions to reach that maximum span. The results of this research show that there is a optimal operating frequency, heat transfer fluid flow rate, flow duration, and displaced volume ratio in an AMR system. By optimizing these parameters the refrigeration performance increased by 24%. It is expected that such optimization will permit the design of a more efficient magnetic refrigeration system.

## I. INTRODUCTION

In contrast with conventional vapor-compression refrigerator systems that work based on compression and evaporation of gas, magnetic refrigeration systems work based on magnetizing and demagnetizing a magnetic material (refrigerant). Magnetic refrigeration exploits a property of magnetic materials called the magnetocaloric effect (MCE): the temperature of ferromagnetic materials is observed to rise upon the application of a magnetic field and fall upon its removal. When a material is magnetized, its magnetic moments are aligned, leading to a reduction in its magnetic entropy. If this process is done adiabatically and reversibly, the total entropy is constant. Thus, a reduction in magnetic entropy is compensated by an increase in lattice entropy resulting in a temperature increase. MCE can be defined as adiabatic temperature change due to magnetization or demagnetization, or, alternatively, as an isothermal magnetic entropy change.

Magnetic refrigeration systems have many advantages compared to conventional refrigerators. They do not use hazardous and ozone-depleting chemicals, greenhouse gases, and are an environmentally friendly cooling technology. Magnetic refrigeration can be more energy efficient than convectional refrigerators. The cooling efficiency of magnetic refrigerators working with gadolinium has been shown to reach a theoretical limit of 60%, while the cooling efficiency of the best gas-compression refrigerators is approximately 40% [1]-[4].

The AMR cycle uses a heat transfer fluid to transport the heat generated or absorbed from MCE in the refrigerant to the hot and cold ends of an AMR device. The AMR cycle consists of four processes: magnetization, hot blowing, demagnetization, and cold blowing. During magnetization, the temperature of the refrigerant increases, then fluid is pumped from the cold end to the hot end in order to absorb the magnetic work (hot blowing). The regenerator is then demagnetized, causing a decrease in temperature, and a cooling load is absorbed from the refrigerant by pumping fluid from the hot end across the regenerator toward the cold end (cold blowing). The four processes need not be discrete, and the fluid flow may coincide with magnetization and demagnetization depending on a system's design.

With the push for the commercialization of the magnetic refrigerator, it is vitally important to evaluate AMR performance. In recent years several AMR experimental systems have been designed and tested [5]-[13]. Most of these focused on the evaluation of the AMR performance by studying different refrigerant materials and multistage systems. Although there are a few researches reporting the effect of a system's parameters on the performance of AMR theoretically, through numerical simulations and modeling [14]-[18], they did not perform any optimization experimentally on their system's parameters.

Three aspects influence the maximum attainable temperature span seen in an AMR device: System, Refrigerant Material, and Thermofluid. Any improvement or optimization done in these categories can result in a higher temperature span seen between the hot and the cold ends of an AMR. The variables for the thermofluid are: type, viscosity, heat conductivity, specific heat capacity, and the tuning of heat transfer fluid and its operating conditions such as frequency, flow rate, flow duration, displaced volume, pressure, etc.

In this work, refrigerant porosity, shape and type, field intensity and distribution, and heat transfer fluid type are fixed parameters, while the experimental focus is on understanding how tuning the heat transfer medium and its operating conditions affect AMR cooling performance.

## II. EXPERIMENTAL SETUP

The schematic design of the AMR system is shown in FIG. 1. The AMR system consists of four main components: A rotary two-core nested permanent magnet cylinder produces a variable field up to 1.3 T, a core glass tube housing the refrigerant bed in the middle of it, six temperature sensors in different distances from the bed, and a reciprocative pump to move the heat transfer fluid through the system between magnetic cycles in order to absorb or release the generated heat by the refrigerant during (de)magnetization.

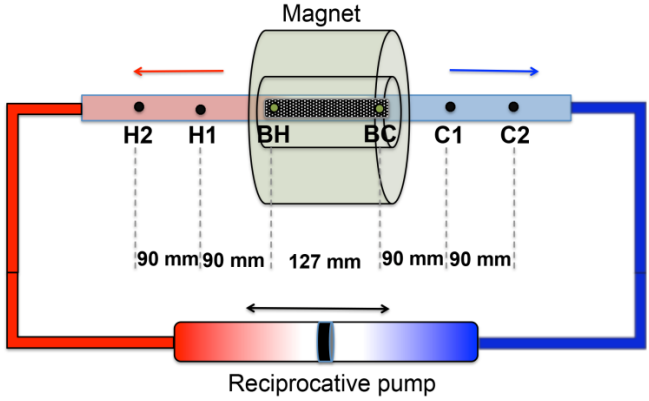

**FIG. 1.** Schematic diagram of the experimental apparatus showing six sensor locations.

Temperature sensors are mounted on holder rods made of Polylactide (PLA) material produced by a 3D printer. Thermocouple readings are disturbed by the proximity of a varying magnetic field [19]. Because sensors BH and BC are exposed to the magnetic field, two Cernox RTD sensors from Lakeshore Cryotronics are used to carry out the measurements at these locations due to their low magnetic field-induced errors. The other sensors (H1, H2, C1, and C2) are outside of magnetic field range; therefore thermocouples were placed at these locations. A picture of the AMR system and the core of the experimental setup are shown in FIG. 2. The volume of the bed is 48cm$^3$ in which 2/3 is filled by gadolinium turnings and 1/3 by heat transfer fluid. In many of the previous studies [5]-[7], [11], the heat transfer fluid was water. However, based on the fact that gadolinium is electropositive and reacts with water to form gadolinium hydroxide and hydrogen gas, it is not a good heat transfer medium candidate. Even though there are corrosion inhibitors that can be added to water to prevent this chemical reaction, silicone oil has additional advantage as its specific heat is almost one fourth of that of water which makes it more suitable since less energy per unit weight is required to raise its temperature by 1 K. Hence, to avoid this chemical reaction and to improve performance, silicone oil is used as the heat transfer medium.

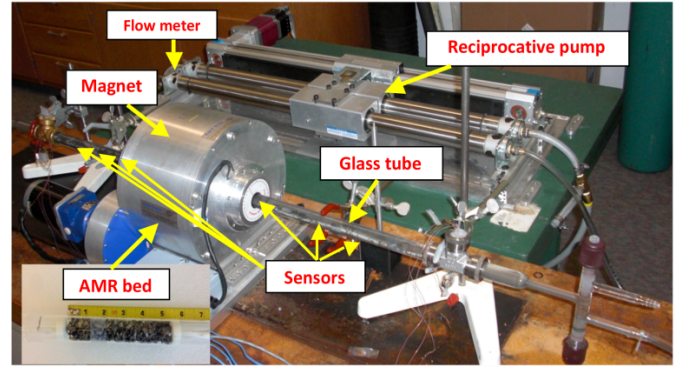

**FIG. 2.** Photography of the AMR device. The refrigerant is located inside a cylindrical permanent magnet assembly.

## III. RESULTS AND DISCUSSION

A number of tests were performed on 140 grams of gadolinium turnings. The temperature span (ΔT) between the hot and cold side of the AMR is measured at multiple operation conditions (different operating frequencies, fluid flow durations, and flow rates). Experimental parameters are presented in Table 1.

**TABLE 1**
Experimental parameters

| Experimental Parameters | Unit | Value |
|---|---|---|
| Operating frequency | Hz | 0.09 – 0.25 |
| Ramp-up & ramp-down time | s | 1.5, 2.6 |
| (de)magnetization time per cycle | s | 0.5 – 4 |
| Gadolinium turnings mass | g | 140 |
| Gadolinium purity | | 99.5% |
| Heat transfer fluid | | Silicone oil |
| Transfer fluid specific heat | J.kg$^{-1}$.K$^{-1}$ | 1370 |
| Fluid flow duration | s | 0.05 – 3 |
| Fluid flow rate | ml.s$^{-1}$ | 1.2 – 86.7 |
| Ambient temperature | C | 21 |
| Bed length | cm | 12.7 |
| Bed radius | cm | 1.1 |

The utilization factor ($\varphi$) is defined as the ratio between the thermal capacity rate of the fluid and that of the solid [7], [10], [16], [17],

$$\varphi = \frac{\dot{m} C_p \tau_{blow}}{M C_H (\mu_0 H = 0)} \quad (1)$$

where $\dot{m}$ is the pumped mass flow rate, $C_p$ is the specific heat capacity of the working fluid, $\tau_{blow}$ is the time period for fluid flow, $M$ is the mass of magnetic material, and $C_H$ is the specific heat of the refrigerant at $\mu_0 H = 0$ T.

In this research a new concept, displaced volume ratio (DVR) is defined as the ratio of the fluid volume that flows through the AMR bed to the volume of the AMR bed occupied by the fluid in each cycle,

$$DVR = \frac{\dot{V} \tau_{blow}}{V_{bed}} \quad (2)$$

where $\dot{V}$ is the fluid volume flow rate (ml.s$^{-1}$), $\tau_{blow}$ is the time period for fluid flow, and $V_{bed}$ is the volume of AMR occupied by fluid. The utilization factor is dependent on the mass and the specific heat of the refrigerant used in the system, whereas the concept of DVR is completely independent of the type and properties of the refrigerant and is solely dependent on the heat transfer medium movement parameters.

Two different setups for each AMR cycle have been introduced in this study. Figure 3 shows these setups as well as one cycle of the experiment. Regions A, B, C, and D are ramp-up, magnetization, ramp-down, and demagnetization, respectively. In both setups the flow rate is kept constant in intervals B and D. In Setup 1 the flow rate for each frequency and DVR is calculated as

$$FlowRate = \frac{V_{BF}}{t_{mag}} \times DVR = \frac{V_{BF}}{\frac{1}{2f} - t_R} \times DVR \qquad (3)$$

where $V_{BF}$ is bed volume occupied by fluid, $t_{mag} = t_{demag}$ is the magnetization or demagnetization time, $f$ is the operating frequency, and $t_R$ is ramp-up or ramp-down duration.

In Setup 2, for each frequency the flow rate is kept constant at its maximum amount and the fluid flow duration varies to generate the desired DVRs. In this setup, right after the beginning of regions B and D there is a delay time ($t_w$) before fluid flow starts. As DVR increases the delay time decreases. During the delay time, there is no fluid movement through the bed. It should be noted that there is no delay time in Setup 1 in which fluid flow starts immediately at the beginning of regions B and D.

Each operating frequency is obtained by changing the duration of regions A, B, C and D. The maximum operating frequency for the system is determined with respect to the maximum allowable fluid flow rate in the apparatus before the seals fail due to excess pressure build up.

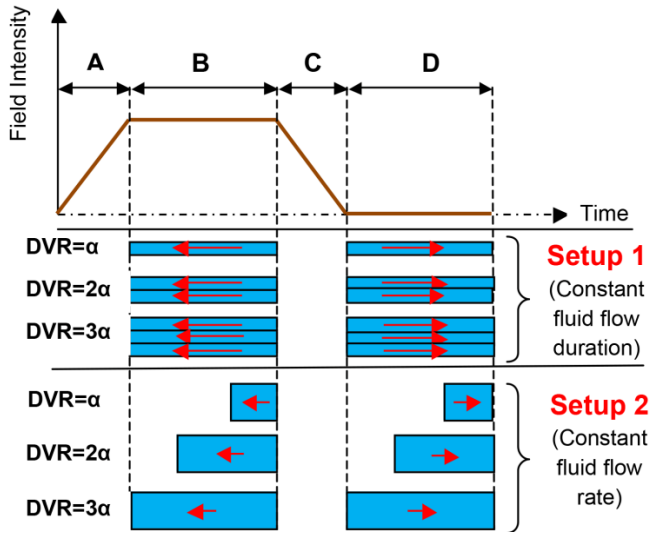

**FIG. 3.** Configuration of Setup 1 and Setup 2 for one cycle of the experiment.

Figure 4 and 5 show the temperature difference between BH and BC sensors versus DVR for different operating frequencies in the two experimental setups respectively. If the value of DVR is too small, only a part of the energy generated by the magnetocaloric effect can be utilized. In addition, too big of a DVR causes the fluid from the hot end of the AMR to flow into the cold end and vice versa, causing a decrease in the ΔT at these locations. The quantity "temperature span" considered herein refers to a single-stage refrigeration cycle and is a property of the refrigerant material rather than of the cooling device. The later can be multistage, with much wider temperature spans [10], [14], [20].

Figure 4 shows ΔT as a function of DVR in Setup 1. There is a DVR of 0.625 for each operation frequency in which the maximum ΔT occurs. Increasing the frequency causes a decrease in the maximum temperature change. This is because at higher operating frequencies there isn't sufficient time to transfer the amount of energy generated by the magnetocaloric effect. In contrast, lower operating frequencies (longer fluid flow durations) provide sufficient time for heat to be transferred from the refrigerant to the fluid (and vice versa) in any cycle, resulting in a higher ΔT. At this optimum DVR, a 24% increase in the performance is achieved by changing the operating frequency.

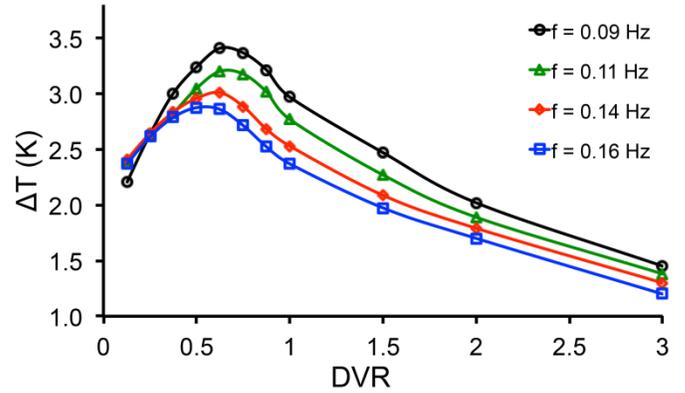

**FIG. 4.** Temperature span ($ΔT_{BH-BC}$) between BH and BC sensors as a function of DVR for different operating frequencies in Setup 1.

The temperature span as a fucntion of DVR in Setup 2 is shown in FIG. 5. The result is that the maximum ΔT occurs at a DVR of 0.3 for all operating frequencies. In Setup 2, unlike Setup 1, as operating frequency increases, ΔT increases. In Setup 2, due to a varying frequency and based on the presence of a delay time for heat exchange between the refrigerant and the fluid, the ΔT is affected by heat losses to the surroundings. As the operating frequency increases, the heat loss during one period decreases, hence ΔT increases. At this optimum DVR, the performance is increase by 22% when varying the operating frequency.

Nonetheless as the operating frequency increases, the delay time diminishes and eventually after some higher DVR the flow occupies the entire region (there won't be a delay time anymore), therefore the two setup configurations exhibit the same behavior (as seen in FIG. 3 at DVR=3α). For instance in FIG. 5 the ΔT curves for frequencies 0.09 Hz and 0.11 Hz cross each others at DVR=2. After this point the delay time for f=0.11 Hz is reduced to zero, therefore Setup 2 behaves like Setup 1 and the ΔT curve for f=0.11 Hz falls below that of f=0.09 Hz. A similar trend occures for the other frequencies at different DVRs and these can be seen in FIG. 5.

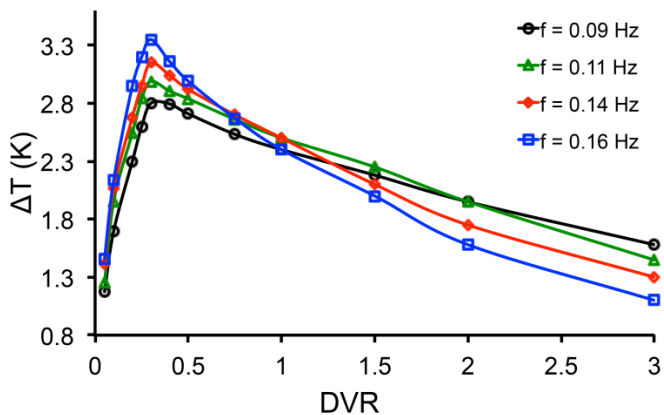

**FIG. 5.** Temperature span ($\Delta T_{BH-BC}$) between BH and BC sensors as a function of DVR for different operating frequencies in Setup 2.

Figure 6 shows the temperature span as a function of frequency for different DVRs in Setup 2. Result shows the ΔT increases as the operation frequency increases and peaks at an optimal frequency of 0.2 Hz. It is worth noting that the optimal frequency is related to the geometry of the refrigerant used in the AMR bed. At this optimum frequency, a 20% increase in the performance is achieved by varying the DVR.

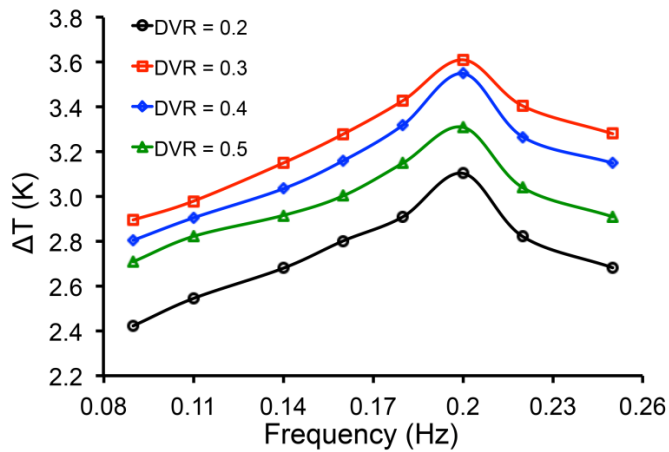

**FIG. 6.** Temperature span as a function of operating frequency for different DVRs in Setup 2.

Figure 7 shows the temperature change at different sensor locations as a function of DVR for the operation frequency of 0.09 Hz. This figure illustrates that the maximum ΔT at various sensor locations occurs at different DVRs. For instance, at locations BH-BC (closest distance to the bed) the peak is reached at DVR=0.3, at locations H1-C1 the peak occurs at DVR=2, and at locations H2-C2 (farthest distance to the bed) the peak is yet to be observed at some DVR higher than 3. Similar trends take place for other tested operating frequencies. This is due to a need for pumping more fluid in order for it to reach sensors which are located further from the AMR bed. This result implies that the performance of an AMR system is dependent on the cooling load placement and its distance from the bed.

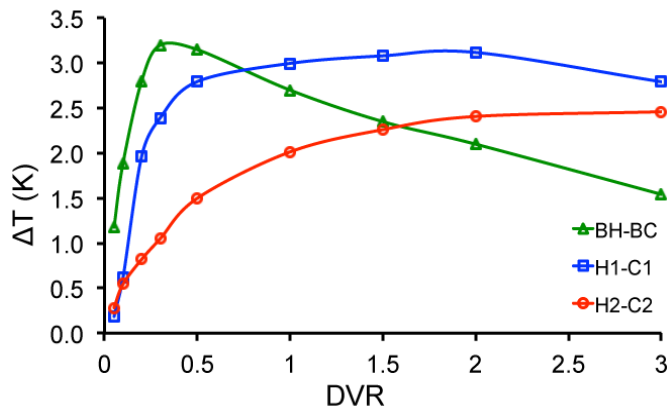

**FIG. 7.** Temperature span at different sensor locations as a function of DVR for a frequency of 0.09 Hz in Setup 2. Sensor locations are shown in FIG. 1. Sensors BH-BC are placed nearest to the bed and sensors H2-C2 are farthest from the bed.

Figure 8 shows ΔT as a function of flow rate in Setup 1. Result shows that different optimal values of the fluid flow rate can be identified at a given frequency with respect to ΔT in order to maximize the performance of the AMR.

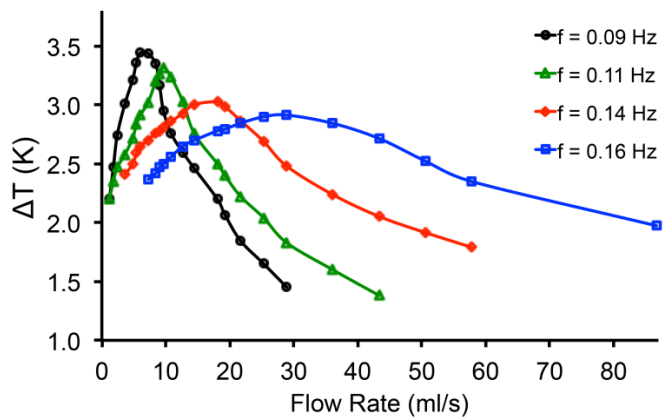

**FIG. 8.** Temperature span as a function of flow rate for different operating frequencies in Setup 1.

## IV. CONCLUSION

In this paper, several significant system parameters that affect the cooling performance of an AMR system have been studied. A single stage AMR test bed using a stationary permanent magnet assembly allowing control of a number of experimental parameters (such as operating frequency, heat transfer medium flow rate, flow duration, and displaced volume ratio) has been designed and evaluated. Two AMR setup configurations were introduced and experiments were performed at different operating conditions.

The results show that the configuration in Setup 1 is more suitable for systems that are operating with lower frequencies, whereas the configuration in Setup 2 is more appropriate for systems that are operating with higher frequencies. It is shown that the performance of an AMR system is dependent on the cooling load placement and its distance from the bed. There were optimal DVRs of 0.625 and 0.3 in Setup 1 and Setup 2, respectively. The optimum operating frequency of the system in Setup 2 was 0.2 Hz in which the highest ΔT occurred at any DVR. Lastly, an optimal value of heat transfer fluid flow rate was identified for any operating frequency. The results of this work demonstrate a significant increase of performance in magnetic refrigeration system (about 24%) by varying and optimizing the AMR parameters. It is expected that such optimization will permit the design of a more efficient commercially viable magnetic refrigeration system.

## ACKNOWLEDGMENTS

The authors would like to thank Mr. James Nunez for his help in setting up the experimental apparatus.